# The stress field and energy of screw dislocation in smectic-A liquid crystals and on the mistakes of the classical solution

Tian You Fan[1] and Xian Fang Li[2]

1 School of Physics, Beijing Institute of Technology

Beijing 100081, China

2 School of Civil Engineering, Central South University, Changsha, Hunan 410075, China

**Abstract** The mistakes of the classical solution of screw dislocations in smectic-A liquid crystals are pointed out. This reveals a serious problem of the well-known theory, which may be named de Gennes-Kleman-Pershan paradox, for many decades in the scientific community of liquid crystal study. The correct solution is given in this letter in terms of simplest, elementary and straightforward solution method. Furthermore, the stress filed and energy of dislocation are discussed in detail. The present article gives a corrected stress field and dislocation energy as well.

PACS number: 61.30.-v, 61.30.Dk , 61.30.Jf

Key words: liquid crystals, dislocation, exact solution

## Introduction

It is well-known that liquid crystals and quasicrystals are the important phases in condensed matter, and fascinating from a fundamental point of view. The liquid crystals as an intermediate phase between isotropic liquids and solids are also significant due to a number of current and potential applications. The mechanical properties of the phase are very interesting, and have extensively been studied, cf. de Gennes and Prost [1], Kleman [2], Kleman and Oswald [3], Oswald and Pieranski [4] etc. But the need of basic research and engineering applications requires promoting the study further, especially those related to defects, three-dimensional elasticity, plasticity and dynamics. Among the defects, dislocations and focal conics have been investigated, see e.g. [1-4] and Landau and Lifshitz [5], Fujii et al [6] etc. Recently, Brostow et al [7] have carried out numerical simulation on crack formation and propagation in polymer liquid crystals. Liquid crystals including smectic ones can be classified as monomer liquid crystals (MLCs), irrespective of the fact whether they can or cannot polymerize and polymer liquid crystals (PLCs). That classification is due to Samuski [8] and has been used (see e.g. [9, 10]). The present model applies to MLCs, in particular smectic-A phase. The study on dislocations in smectic-A liquid crystals has a long history, at least several decades so far. In the work of the solution of screw dislocation of smectic-A liquid crystals, there is long-standing puzzle. Though Kleman [2] and Pershan [11] independently found the well-known solution of screw dislocation in smectic-A liquid crystals

$$u = \frac{b\theta}{2\pi},\qquad(1)$$

in which $u$ is the displacement along the direction of $z$-axis, normal to the layers of smectics A, $b$ the magnitude of the Burgers vector of the screw dislocation, and $\theta$ the polar angle of the position vector in the $xy$–plane, it belongs to a classical work in the discipline and presents important sense to later study. The original results given by Kleman [2] and by Pershan [11] are based the de Gennes' equation of elasticity of smectic-A liquid crystals (i.e., equation (5) in the following), one of the most famous and important results in the liquid crystal study, see e.g. Refs [1,2,4], The result has been widely cited by many other authorized monographs, e.g. [5] etc. Though certain researchers (e.g. Pleiner[12]) criticized the de Gennes-Kleman-Pershan's solution, he still confirmed that solution (1) holds in the region outside the dislocation core. Unfortunately the idea has been widely accepted. A further discussion on the solution is necessary. The above solution is a solution out of the core of the dislocation. Kralj and Sluckin [13,14] studied the core structure of a screw dislocation in smectic-A liquid crystals, which are very interesting and important results. The core structure naturally influences the solution out of the core. But at present our attention is focused on the solution out of the core. If we can correctly explore the solution, this may help us to reveal the core structure.

**I Basis of discussion**

There have been many discussions from different angles so far, e.g. the magnetism analogue [1,11], the differential geometry [15], the dynamics [16], the structure of dislocation core [13,14,17], etc. Although these discussions from different points of view are beneficial, this leads to some difficulty to the readers. To ensure the discussion to be arrived in exact agreement, a brief basis of elasticity of smectic-A liquid crystals is outlined.

The elasticity of smectic-A liquid crystals was established by de Gennes [1] first, and an outstandingly simplified formulation was presented by Landau and Lifshitz [5]. Though the theory and formulation are well known, it is very beneficial to recall them once again for the discussion.

In macroscope, within the framework of continuum liquid crystals, Landau-Ginzburg-de Gennes free energy of the deformation of smectic-A liquid crystals reads [5]

$$F_d = F - F_0(T) = \frac{1}{2}(A/\rho_0)(\rho-\rho_0)^2 + C(\rho-\rho_0)\frac{\partial u}{\partial z} + \frac{1}{2}B\rho_0\frac{\partial^2 u}{\partial z^2} + \frac{1}{2}K_1\left(\nabla^2 u\right)^2$$
$$= \frac{1}{2}\rho_0 B'\frac{\partial^2 u}{\partial z^2} + \frac{1}{2}K_1\left(\nabla^2 u\right)^2$$

where

$$\nabla^2 = \frac{\partial^2}{\partial x^2} + \frac{\partial^2}{\partial y^2}, \rho - \rho_0 = -\rho_0 m\frac{\partial u}{\partial z}, m = \rho_0\frac{C}{A}, B' = B - \frac{C^2}{A}$$

in which $\rho_0 B'$ and $K_1$ are Young's modulus and splay modulus,

respectively, and there are no deformation of twisting and bending (i.e., $K_2 = 0$ and $K_3 = 0$).

Now consider a smectic-A liquid crystal with a screw dislocation along the direction of $z$-axis. The physical model can be stated by the following boundary conditions:

$$\left. \begin{array}{l} (x^2 + y^2)^{1/2} \to \infty: \quad \sigma_{ij} = 0 \\ \sigma_{zz}(x,0) = 0 \\ \int_\Gamma du = b \end{array} \right\} \quad (2)$$

in which $\sigma_{ij}$ denotes the stress tensor, and $\Gamma$ denotes a closed contour enclosing the dislocation core. The solution of stress field induced by a screw dislocation needs to introduce relevant basic equations.

According to de Gennes and Prost [1] or Landau and Lifshitz [5] the stress tensor for smectic-A liquid crystals (exactly speaking for smectics-A) is

$$\left. \begin{array}{l} \sigma_{xx} = \sigma_{yy} = K_1 \nabla^2 \dfrac{\partial u}{\partial z} \\ \sigma_{zz} = \rho_0 B' \dfrac{\partial u}{\partial z} \\ \sigma_{zx} = \sigma_{xz} = -K_1 \nabla^2 \dfrac{\partial u}{\partial x} \\ \sigma_{zy} = \sigma_{yz} = -K_1 \nabla^2 \dfrac{\partial u}{\partial y} \\ \sigma_{xy} = \sigma_{yx} = 0 \end{array} \right\} \quad (3)$$

in which $u$ is the displacement component in the $z$-axis, and $B' = B - C^2/A$, $A, B, C$ and $K_1$ are materials constants, cf. [1,5]. Note that the constant $B'$ describes material modulus relating to deformation arising from displacements, while $K_1$ describes material modulus relating to deformation arising from curvature. It is evident in the stress tensor (3) that only the symmetry part, according to the terminology of Landau and Lifshitz [5], is considered. The viscous part is omitted here by following the methodology adopted in the community of liquid crystals.

Substituting equations (3) into the equilibrium equations

$$\frac{\partial \sigma_{ij}}{\partial x_j} = 0 \qquad (4)$$

yields a final governing equation

$$\rho_0 B' \frac{\partial^2 u}{\partial z^2} - K_1 \nabla^2 \nabla^2 u = 0 \qquad (5)$$

which was derived from the energy minimization by de Gennes [1] where $\nabla^2 = \frac{\partial^2}{\partial x^2} + \frac{\partial^2}{\partial y^2}$. Thus, the solution of a screw dislocation is reduced to solving equation (5) subjected to the boundary conditions (2), which can be called the boundary value problem "(5), (2)".

Because the problem is assumed uniform along the $z-$ axis, then any field variable is independent of $z$, i.e.,

$$\frac{\partial ()}{\partial z} = 0 \qquad (6)$$

where () represents any field functions. So equation (5) is reduced to

$$\nabla^2 \nabla^2 u = 0 \qquad (7)$$

In this case the boundary value problem "(5),(2)" is replaced by the boundary value problem "(7),(2)".

Equations (2)-(7) are the basis in common in the discussion. If the discussion follows the basis, some confusion in methodology and logics can be avoided.

**II The simplest and most direct solution way**

To let major readers easily understand the discussion, we suggest to take the simplest, elementary and straight solution method to solve the boundary value problem "(7),(2)", and need not to use the magnetism analogue, or Fourier transform, or Green function etc. Some references made the problem to be complexity by using complicated mathematical methods. In contrast, we take an alternative way, in which the analysis is extremely simplified.

Introducing polar coordinate system $(r,\theta)$, equation (7) is rewritten as

$$\left(\frac{\partial^2}{\partial r^2}+\frac{1}{r}\frac{\partial}{\partial r}+\frac{1}{r^2}\frac{\partial^2}{\partial \theta^2}\right)\left(\frac{\partial^2}{\partial r^2}+\frac{1}{r}\frac{\partial}{\partial r}+\frac{1}{r^2}\frac{\partial^2}{\partial \theta^2}\right)u(r,\theta)=0 \qquad (7')$$

A suitable solution of equation (7') through the variable separation method, i.e., $u(r,\theta)=f(r)\Theta(\theta)$, takes the following form

$$u=\frac{b}{2\pi}\left[\left(Dr^2+Er^2\ln r+F+G\ln r\right)\theta+\left(D_1 r+E_1 r\ln r\right)\theta\sin\theta+\left(F_1 r+G_1 r\ln r\right)\theta\cos\theta\right]$$

where we have neglected the terms which are independent of the solution of the dislocation. In other words, those terms that only cause an increment in angle when going a circuit around the dislocation core are retained. Making use of the condition (2), we find that the parts related to $D$, $E$ and $G$ give rise to an increment dependent on $r$ when running around the dislocation core. After removing the terms related to $D$, $E$ and $G$, a suitable solution further takes the following form

$$u = \frac{b}{2\pi}\left[F + (D_1 r + E_1 r \ln r)\sin\theta + (F_1 r + G_1 r \ln r)\cos\theta\right]\theta \tag{8}$$

in which the unknown constants $E_1$ and $G_1$ vanish by considering stress continuity. Consequently, solution (8) at last becomes

$$u = \frac{b}{2\pi}\left[F + D_1 r \sin\theta + F_1 r \cos\theta\right]\theta \tag{8'}$$

Furthermore, due to conditions (2) one can determine $F=1$ and $F_1=0$. However, the unknown constant $D_1$ still cannot be determined. To determine the value of $D_1$, let us give the stress field. This can be done by substituting (8') into (3), yielding

$$\left.\begin{array}{l}\sigma_{zx} = \dfrac{b}{2\pi}\dfrac{2K_1 D_1(x^2 - y^2)}{(x^2 + y^2)^2} \\[2ex] \sigma_{zy} = \dfrac{b}{2\pi}\dfrac{4K_1 D_1 yx}{(x^2 + y^2)^2}\end{array}\right\} \tag{9}$$

In monograph [18] we developed the mathematical theory of dislocations in quasicrystals, in which the problems of crystals are naturally included (because if the phason field is absent, the quasicrystals are reduced to crystals). The theory demonstrated that the higher partial differential equations describing dislocations need appropriate additional boundary conditions except the dislocation condition; otherwise the boundary value problem will not be well-defined. This is valid for boundary value problem "(7), (2)" too. It is sufficient to determine the unknown constants $F$ and $D_1$ with the aid of two conditions in expressions (2) except the condition at infinity. Because $\sigma_{zz}(x,0)=0$ is automatically satisfied, we must now search an additional condition determining the third constant. We here use the minimization of dislocation energy, i.e.,

$$\frac{\partial U}{\partial D_1} = 0 \tag{10}$$

where the energy will be given in the following (i.e., (equations (17-19)). From this condition we have

$$D_1 = \frac{-\frac{8}{3}\pi\alpha}{(R_0 + r_0)\left(\frac{\pi}{4}\alpha\beta + \frac{b^2 K_1}{8\pi}\right)\ln\frac{R_0}{r_0} + \frac{\pi}{320}\alpha\gamma(R_0 - r_0)} \tag{11}$$

in which

$$\left.\begin{array}{l} \alpha = \left(\dfrac{b}{2\pi}\right)^4 \rho_0 B' \\[6pt] \beta = 2 + \dfrac{32\pi^2}{3} \\[6pt] \gamma = 75 - 160\pi^2 + 256\pi^4 \end{array}\right\} \tag{12}$$

From equations (8) and (1), one can find the solution given by Kleman [2] and Pershan [11] is only one of terms of the present solution. In other words, the classical solution is the zero-order approximation of the present solution. In particular, the classical solution does not induce any stress, or the dislocation causes stress-free state, while according to our solution, the stress field exits, and exhibits a square singularity near the dislocation core. This singularity is also different from the stress field induced by a screw dislocation in conventional crystals. For the latter, the stress field has a $r^{-1}$ singularity, rather than $r^{-2}$ singularity. In addition, this singularity is also different from the square-root singularity near a crack tip.

### III Mathematical mistakes of the classical solution

In solid crystal elasticity (or classical elasticity) the screw dislocation problem is formulated by

$$(x^2 + y^2)^{1/2} \to \infty : \sigma_{ij}^{(c)} = 0 \atop \int_\Gamma du^{(c)} = b^{(c)} \bigg\} \quad (13)$$

and

$$\nabla^2 u^{(c)} = 0 \quad (14)$$

in which

$$\sigma_{zx}^{(c)} = \sigma_{xz}^{(c)} = \mu \frac{\partial u^{(c)}}{\partial x} \atop \sigma_{zy}^{(c)} = \sigma_{yz}^{(c)} = \mu \frac{\partial u^{(c)}}{\partial y} \bigg\} \quad (15)$$

where the superscript (c) represents crystal, $\mu$ the shear modulus of the crystal.

The solution (1) is only the solution of boundary value problem "(14), (13)". According to the theory of partial differential equations or mathematical theory of elasticity [19, 20], the solution (1) of boundary value problem "(14), (13)" cannot be the solution of boundary value problem "(7), (2)" at the same time. The problem cannot be solved by the so-called smallest surface concept, provided equations (3) and (5) are invalid. We believe that equations (3) and (5) (and its reduced form (7)) are valid, and the stress field induced by a single straight screw dislocation along the $z$–axis can be determined by solving equation (7) subjected boundary condition (2). The solution (8) is the unique solution of boundary value problem "(7), (2)".

**IV The physical mistakes of the classical solution**

The solution (1) leads to some physical mistakes too. This can be viewed in the following.

1) It leads to zero stress field. Substituting solution (1) into equations (3) leads to

$$\sigma_{ij} = 0, \quad i, j = 1, 2, 3 \tag{16}$$

This is completely wrong physically.

2) It leads to wrong energy formulas. The energy induced by the dislocation is one of important aspects of the problem. On the calculation of energy induced by the dislocation, there are many contradictions between de Gennes [1], Oswald and Pieranski [4], Kleman [15] and Pleiner [17], even if in the monograph [4] there are logic contradiction itself. This shows the difficulty of the problem. According to our understanding, the point of view of Kleman in [15] is correct, though his calculation is not complete, in which there are some mistakes because he used the solution (1). Adopting the point of view of Kleman, the energy consists of three parts: 1) arising from splay, 2) arising from bulk deformation, 3) corresponding to the dislocation core energy, i.e.,

$$\begin{cases} U_1 = \iint_A \frac{1}{2} K_1 \left( \nabla^2 u \right)^2 dxdy \\ U_2 = \iint_A \frac{1}{2} \rho_0 B' \left[ \left( \frac{\partial u}{\partial x} \right)^2 + \left( \frac{\partial u}{\partial y} \right)^2 \right]^2 dxdy \\ U_3 = \frac{1}{2} \iint_\Omega (\sigma_{zx} \frac{\partial u}{\partial x} + \sigma_{zy} \frac{\partial u}{\partial y}) dxdy \\ \phantom{U_3} = \frac{1}{2} \int_{r_0}^{R_0} \int_0^{2\pi} r(\sigma_{zx} \frac{\partial u}{\partial x} + \sigma_{zy} \frac{\partial u}{\partial y}) drd\theta \end{cases} \qquad (17)$$

$$U = U_1 + U_2 + U_3 \qquad (18)$$

where $A$ represents the integration domain—the total $xy$–plane, and $R_0$ and $r_0$ the conventional outer and inner radii in calculating dislocation energy. If substituting solution (1) into the energy equation (18), one can obtain the wrong results only.

According to the Landau-Ginzburg-de Gennes free energy, the first part of equation (18) always vanishes. When substituting (8) into the first, second and third part, one gets $U_1$=0, and

$$U_2 = \frac{\pi}{8} \left( \frac{b}{2\pi} \right)^4 \rho_0 B' \left[ \frac{1}{r_0^2} - \frac{1}{R_0^2} \right] + \frac{\pi}{8} \left( \frac{b}{2\pi} \right)^4 \rho_0 B' D_1^2 \left[ 2 + \frac{32\pi^2}{3} \right] \ln \frac{R_0}{r_0}$$
$$+ \frac{D_1}{240} \frac{\pi}{8} \left( \frac{b}{2\pi} \right)^4 \rho_0 B' \left[ \frac{5120}{R_0 + r_0} + 3D_1 \left( 75 - 160\pi^2 + 256\pi^4 \right) \right] \left( R_0^2 - r_0^2 \right) \qquad (19)$$
$$U_3 = \frac{b^2 K_1 D_1^2}{16\pi} \ln \frac{R_0}{r_0}$$

The solution (1) cannot obtain these energy expressions. The solution (1) does not hold for smectic-A liquid crystals, even if under the condition of continuum model. The invalidity presents not only in the region inside the core of dislocation, but also in the region outside the core of dislocation.

## V Meaning of present solution

The solution (8) overcomes the mistakes of well-known classical solution (1) mentioned above, and from which we obtain some meaningful and useful results. For example, we can evaluate the dislocation core energy of smectic-A, and find that the dislocation energy is correlated to both Young's modulus and splay modulus. Another finding is that the stresses obtained from solution (8) exhibit

$$\sigma_{xz}, \sigma_{yz} \sim r^{-2}, \quad r \to 0 \tag{20}$$

This presents a different singularity as that in quasicrystals [18] apart from crystals.

## VI Conclusion and discussion

In summary, by introducing the fundamental physical framework as the basis of the discussion and based on the existence and uniqueness principle of partial differential equations, we presented the simplest, elementary and straight solution method to demonstrate the mistakes of the well-known solution (1) spreading almost 4 decades and to derive the correct solution. This proves that the classical solution is only the zero-order approximation of the present correct solution. The discussion may provide an effort to solve a long-standing puzzle in studying screw dislocations in smectic-A liquid crystals. Due to the difficulty of the problem we cannot say that this paper solves the all aspects of the long-standing puzzle. In addition the stress singularity of dislocation in the phase is different from that in crystals and quasicrystals, and this may be a meaningful finding. The discussion on the dislocation core structure is left for further study, and a correct solution out of the core may be a basis for discussing core structure.

**Acknowledgement**   The authors thank Professor Pleiner H for the references [12], [16], [17] available.